# Formation of Uniform Crystal and Reduction of Electrical Variation in HfZrO$_2$ Ferroelectric Memory by Thermal Engineering

Sourav De, *Student Member, IEEE*, Bo-Han Qiu, Md. Aftab Baig, Darsen D. Lu\*, *Senior Member, IEEE*, and Yao-Jen Lee\*

*Abstract*—In this paper we proclaim excellent variation control in Hf$_{0.5}$Zr$_{0.5}$O$_2$ based ferroelectric films obtained by germination of large ferroelectric domain via extended duration of thermal annealing. 10nm thick Hf$_{0.5}$Zr$_{0.5}$O$_2$ based ferroelectric capacitors with TiN as bottom and top electrodes are fabricated and characterized. The duration of rapid thermal annealing (RTA) is varied to observe its effect on crystal formation and device electrical properties at 700°C. The device to device variation in terms of coercive voltage and peak capacitance are reduced from 0.4V to 0.01V and from $2\times10^{-5}$nF/cm$^2$ to $4\times10^{-6}$nF/cm$^2$, respectively, by increasing the RTA duration. High resolution transmission electron micrograph clearly shows large and uniform ferroelectric domains with RTA of 180 seconds. Extended duration of RTA likely allows uniform crystal to form, which mitigates the stochasticity of the distribution of ferroelectric and paraelectric domains, and deterministic switching has been infused. This improvement paves the way for implementing Hf$_{0.5}$Zr$_{0.5}$O$_2$ based deeply scaled devices for memory and steep slope device applications.

## I. Introduction

RECENT progresses in CMOS-compatible and deeply-scalable hafnium zirconium oxide based ferroelectric FETs(Fe-FET) have manifested its potentiality for being used as a steep slope device, low power electronics, memory and neuromorphic applications [1],[2]. Amidst many advantages like low latency, CMOS-compatibility and high endurance the device to device variations infused by erratic distribution of ferroelectric phase in Hf$_{0.5}$Zr$_{0.5}$O$_2$ (HZO) engenders reliability issues in the operation of Fe-FETs [3-12]. This random distribution of ferroelectric-paraelectric phase affects the coercive voltage of a ferroelectric capacitor, which inevitably affects the forward and reverse threshold voltage of Fe-FET devices, leading to variations in program and erase voltage. The impact of this variations on neural network applications has been shown in [10,11,12].

In order to apply ferroelectric FETs for reliable memory array, IC and circuit design the mitigation of these variations are of paramount importance. Although some recent attempts demonstrate various options like higher annealing temperature [13], altering the atomic layer deposition cycle ratio [14] or insertion of additional layers [15] in the dielectric stack to minimize the variations in HZO by increasing the number of grains, we are yet to reach any peroration about the any specific process described in literature to alleviate the random polycrystallinity induced variations. In this letter we discuss about our approach of increasing the annealing time to allow germination of large ferroelectric grain, which is similar to single crystalline phase in HZO. The analysis, showing the impact of annealing duration on the variation of coercive voltage and peak capacitance, corroborates that increasing annealing time leads to reduction of the device to device variations in HZO.

## II. Kinetics of Crystallization in HZO:

It is quite well known from the kinetics of crystallization, that under isothermal condition the nucleation of a phase in a solid with unbounded volume follows the JMAK equation, where the fraction of transformed volume can be given by $x(t) = 1 - e^{-K.t^n}$ [17]. The JMAK theory also says that, any crystal point will be transformed into the state x' from x by time t if and only if the adjacent points are already nucleated and that very particle gets sufficient time to grow into x'. Albeit originally being developed for infinite volume, this concept can be carried out for the phase change in finite volume or thin film solids with possible corrections shown in [16,17,18]. In isothermal condition, the growth rate or the phase transformation probability of any point will be dependent on the time and thickness. The modified JMAK equation for HZO can be written as, $X(t, T_{FE}) = 1 - \int_{-\frac{T_{FE}}{2}}^{\frac{T_{FE}}{2}} exp\{-\emptyset(t,z)\}dz$ [16]. $X(t, T_{FE})$ is the fraction of volume with new phase during thermal treatment. $T_{FE}$ is the thickness of ferroelectric material. The term $-\emptyset(t,z)$ can be related to the probability of a point being transformed to a new phase as it is done in [16]. It is quite evident from these equations that during RTA or isothermal process of phase change any point requires adequate time for nucleation. The nucleation of any point occurs only when the adjacent points are nucleated into new phase. Although, increasing the temperature may increase the number of nucleation sites and increase the number of ferroelectric grains, but RTA with higher temperature damages the HZO film by infusing resistive leakage. In the following section we shall see

This work was jointly supported by the Ministry of Science and Technology (Taiwan) grant MOST-108-2634-F-006-08 and is part of research work by MOST's AI Biomedical Research Center.

S. De, B.H.Qiu, Aftab Baig and D.-D. Lu are with the Department of Electrical Engineering and the Institute of Microelectronics of National Cheng Kung University (NCKU), Tainan, Taiwan R.O.C. (e-mail: darsenlu@mail.ncku.edu.tw)
Y.-J. Lee is with Taiwan Semiconductor Research Institute, Hsinchu, Taiwan R.O.C. (e-mail: yjlee@narlabs.org.tw)

how this time variable plays a pivotal role in the mitigating the variation in HZO based ferroelectric materials.

III. FABRICATION AND MATERIAL CHARACTERIZATION:

HZO film of 10nm thickness is developed on TiN/SiO$_2$/Si(n+) by atomic layer deposition (ALD) method at 250ºC. TEMAHf, and TEMAZr and H$_2$O were used as the precursor for hafnium, zirconium and oxygen respectively. The deposition of ZrO$_2$ was done at first and one cycle of ZrO$_2$ was followed by one cycle of HfO$_2$. TiN was also used as top electrode to minimize the work function mismatch generated asymmetricity [19]. 2.5nm thick SiO$_2$ layer is used at silicon HZO interface to block channel side trapping [20], which shrinks the hysteresis and infuse undesired trapping and de-trapping instigated variation. Both bottom and top electrodes, deposited using physical vapor deposition (PVD) method, are 20nm thick. Post deposition of top TiN electrode the square shaped capacitors of area 10000μm$^2$ were formed by lithography, followed by rapid thermal annealing (RTA) at 700ºC for 30s, 40s, 90s and 180s respectively.

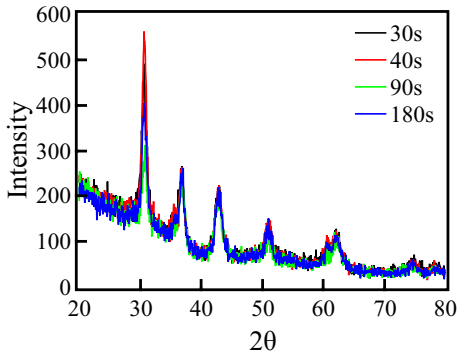

Fig.1. X-ray diffraction analysis of HZO annealed for 30s ,40s, 90s and 180s.

Fig. 1 shows the X-ray diffraction (XRD) analysis of the HZO capacitors annealed for four different time duration. The peaks have been identified according to the data from [21]. The highest peak of cubic, orthorhombic, monoclinic and tetragonal variants of HfO$_2$ and ZrO$_2$ are observed around 30º. Therefore, it is non-viable to identify the existence of any phase or calculate the fractional volume of any phase from the XRD data.

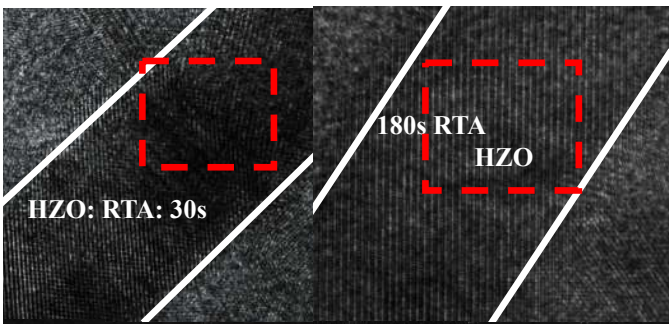

oh showing the dimensions of
EM of capacitors annealed for
ifferent crystal directions, mani
s. While the 180s-annealed samp
n within the entire HR-TEM.

Therefore, we adopted the high-resolution transmission electron micrograph (HR-TEM) technique to characterize the growth of the ferroelectric grain with various annealing time. The HR-TEM image in Fig. 2 shows the crystal is very uniform for the sample annealed for 180s, substantiating near-perfect crystallization. Whereas indistinct grain boundary is extant for HZO films annealed for 30s. This nanoscale material analysis corroborates better crystallization, when the annealing time is increased. In the following sections we shall divulge our investigation on the effect of annealing time on the electrical characteristics of HZO film.

IV. ELECTRICAL CHARACTERIZATION AND RESULTS

The electrical characterization is done on pristine HZO films by *Radiant Precision LC* measurement system at 10kHz using a triangular voltage peak to peak voltage of 6V.

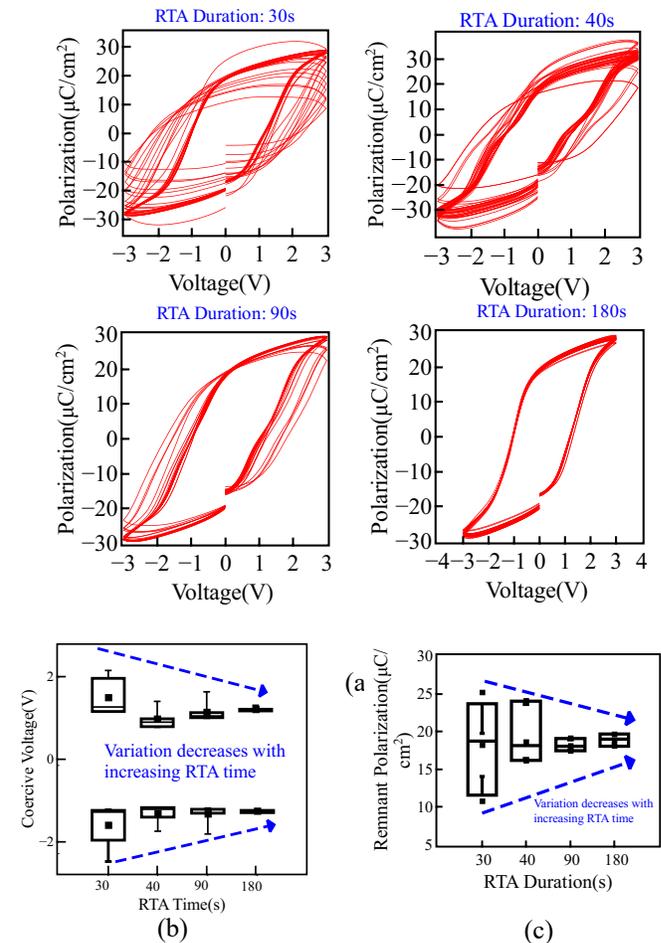

Fig. 3. (a) PV curve of 10nm thick TiN/HZO/TiN/SiO$_2$/Si stack with different annealing time (b) The variation in coercive voltage decreases with increasing RTA duration. (c). Although the maximum remnant polarization decreases, the variation in remnant arization is reduced, leading to more stable characteristics.

3(a) shows the polarization-voltage (P-V) curve of /HZO/TiN/SiO$_2$/Si stack with an active area of 10000μm$^2$ ealed at 700ºC with four different durations. The variation ysis of coercive voltage (Fig. 3(b)) and remnant polarization (Fig. 3(b)) shows that the variation reduces when the annealing time is increased to 180s from 30s. Fig. 3(b) shows the distribution of coercive voltage up to ±2.5V and the coercive voltage distribution gets tapered with increasing RTA duration. Although a high coercive voltage might seem to be alluring option for obtaining wider memory window (MW), the variation associated with this is not desired. We can find the similitude of this trend in Fig. 3(c), where the distribution of remnant polarization also slender with increasing RTA duration. Fig. 4 shows the cumulative distribution along with probability distribution of coercive voltage for four different RTA duration. The HZO film is paraelectric as deposited. During annealing the ferroelectric orthorhombic Pca2$_1$ phase is formed in HZO at the constant annealing temperature. According to crystallization kinetics (Section II), increasing the time of annealing will lead to increase in the fraction of ferroelectric phase over paraelectric phase. The nucleation percentage of ferroelectric phase increases over time until the saturation is reached. The co-existence of paraelectric and ferroelectric phase can be attributed to the incomplete nucleation of ferroelectric phase caused by annealing conducted for inadequate time. P-V characteristics (Fig. 3(a)) indicate that without sufficient annealing time, a large percentage of devices exhibit paraelectric P-V curves. Furthermore, the presence of two peaks in coercive field distribution is a strong indication of incomplete formation of ferroelectric phase during the RTA process (Fig. 4(a)(b)). The peak around ±1V corresponds to ferroelectric domains, whereas the peak around ±2V is related to paraelectric domains. Therefore, two peaks indicate the co-existence of ferroelectric and paraelectric domains. The second peak (around ±2V) gradually diminishes when the duration of annealing is increased, indicating the disappearance of paraelectric cases (Fig.4(d)). It can be observed from Fig. 4 that as we increase the duration of RTA, the second peak around ±2V starts to diminish along with the reduction of standard deviation of forward and reverse coercive field to 30mV and 10mV from 400mV and 440mV, respectively.

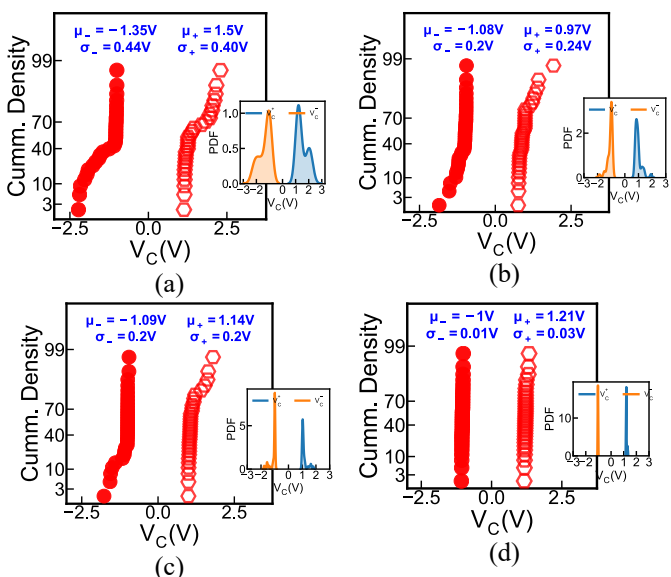

Fig. 4. CDF and PDF (inset) of coercive voltage of fabricated ferroelectric capacitors annealed at (a) 30 seconds (b) 40 seconds (c) 90 seconds (d) 180 seconds.

## V. CONCLUSION

10nm thick HZO based ferroelectric capacitors are fabricated and characterized, showing excellent variation control when the RTA duration is increased to 180 seconds. Such simple physics-based method conveniently reduce device to device variation caused by the coexistence of ferroelectric and paraelectric phases in HZO. This method paves the way for reducing the variability in deeply scaled Fe-FET devices.

## ACKNOWLEDGEMENTS

We are grateful to Taiwan Semiconductor Research Institute for nanofabrication facilities and services, and Dr. Wen-Jay Lee and Nan-Yow Chen of National Center for High-Performance Computing for helpful suggestions on AI computation.